\documentclass[aps, a4paper, preprint, superscriptaddress, preprintnumbers, floatfix, nofootinbib, amsmath, amssymb ]{revtex4}

\usepackage[utf8x]{inputenc}
\usepackage{graphicx}
\usepackage{hyperref}
\usepackage{float}
\usepackage{amsmath}
\usepackage[toc,title,page]{appendix}
\usepackage{setspace}
\usepackage{multirow}
\usepackage{natbib}
\usepackage{subfigure}
\begin{document}
\title{Implication of cosmological upper bound on the validity of golden ratio neutrino mixings under radiative corrections}
\author{Y Monitar Singh}
\email{monitar.phd.phy@manipuruniv.ac.in}
\affiliation{Department of Physics, Manipur University, Imphal-795003, India}
\author{M Shubhakanta Singh}
\email{mshubhakanta@yahoo.com} 
\affiliation{Department of Physics, Manipur University, Imphal-795003, India}
\author{N. Nimai Singh}
\email{nimai03@yahoo.com} 
\affiliation{Department of Physics, Manipur University, Imphal-795003, India}
\affiliation{Research Institute of Science and Technology, Imphal-795003, India}

\begin{abstract}
We study the implication of the most recent cosmological upper bound on the sum of three neutrino masses, on the validity of the golden ratio (GR) neutrino mixings defined at high energy seesaw scale, considering the possibility for generating low energy values of neutrino oscillation parameters through radiative corrections in the minimal supersymmetric standard model (MSSM). The present study is consistent with the most stringent and latest Planck data on cosmological upper bound, $\sum |m_{i}| < 0.12$ eV. For the radiative generation of sin$\theta_{13}$ from an exact form of golden ratio (GR) neutrino mixing matrix defined at high seesaw energy scale, we take opposite CP parity mass eigenvalues ($m_{1},-m_{2},m_{3}$) with a non-zero real value of $m_{3}$, and a larger value of $\tan\beta > 60$ in order to include large effects of radiative corrections in the calculation. The present analysis including the CP violating Dirac phase and SUSY threshold corrections, shows the validity of golden ratio neutrino mixings defined at high seesaw energy scale in the normal hierarchical (NH) model. The numerical analysis with the variations of four parameters viz. $M_{R}$, $m_{s}$, $\tan\beta$ and $\bar{\eta_{b}}$, shows that the best result for the validity is obtained at  $M_{R}=10^{15}$ GeV, $m_{s}=1$ TeV, $\tan\beta=68$ and $\bar{\eta_{b}}=0.01$. However, the analysis based on inverted hierarchical (IH) model does not conform with this latest Planck data on cosmological bound but it still conforms with earlier Planck cosmological upper bound $\sum |m_{i}|< 0.23$ eV, thus indicating possible preference of NH over IH models.

\end{abstract}

\maketitle
\section{Introduction}

The values of neutrino oscillation parameters have been continuously updated with the advancement in the technology of neutrino oscillation experiments  \cite{esteban2020fate,de20212020,mc} and these updated experimental data are also required to compare with the theoretically predicted values. The latest Planck data on the cosmological upper bound on the sum of the three absolute mass eigenvalues given by $\sum |m_{i}|<0.12$ eV \cite{aghanim2020planck}, may be seriously considered while comparing with other neutrino oscillation parameters although there are also a lot of constraints associated with such cosmological probe\cite{tanseri2022updated}. The theoretical predictions of these neutrino oscillation parameters are in general defined at very high energy seesaw scale, and the experimental data on the other hand are defined at low energy scale of the order of $10^{2}$ GeV. In order to make a bridge between these two energy scales, we need a set of renormalisation group equations (RGEs) for quantum radiative corrections \cite{das2005numerical,king2000renormalisation}. We can use two different approaches for running the RGEs from high-energy scale to low-energy scale. In the first approach, the running of RGEs is carried out through the neutrino mass matrix $m_{LL}$ as a whole, and at every energy scale one can extract neutrino masses and mixing angles through the diagonalisation of the neutrino mass matrix calculated at that particular energy scale \cite{chankowski1993renormalization,babu1993renormalization,as,parida1998low,king2001inverted}. In the second approach, the running of RGEs can be carried out directly in terms of neutrino mass eigenvalues and three mixing angles with phases \cite{chankowski2000fixed,brignole2003low,antusch2003running}. In both cases, the RGEs of all the neutrino parameters and the RGEs of various coupling constants are solved simultaneously and both approaches give almost consistent results \cite{das2005numerical}. For the present analysis, we shall use second approach which is more convenient to handle in the numerical analysis of RGEs of neutrino oscillation parameters.
\newline

 Various discrete symmetry groups like $S_{4}$, $A_{4}$, $A_{5}$ etc. which are defined at very high energy scale, can lead to various leptonic mixing matrices such as bi-maximal (BM), tri-bimaximal (TBM) and Golden ratio (GR)\cite{Kajiyama_2007}. All these specific leptonic mixing matrices have their own respective leptonic mixing angles, and two of the mixing angles ($\theta_{23}$ and $\theta_{12}$) are in good agreement with the respective non-zero neutrino mixing angles at low energy scale. In all the above three leptonic mixing matrices, the three leptonic neutrino mixing angles are defined at very high energy scale, with reactor neutrino mixing angle ($\theta_{13}$) equals to zero. The radiative magnification of reactor neutrino mixing angle ($\theta_{13}$) is studied with various leptonic mixing matrices such as BM, TBM and GR \cite{zhang2016viability,pramanick2021radiative}.
 
  GR neutrino mixing pattern has certain advantages over the other two neutrino mixing patterns BM and TBM in the evolution of mixing angles under radiative corrections as solar mixing angle ($\theta_{12}$) is always found to increase with deccrease in energy scale. The generation of right order non-zero value of reactor neutrino mixing angle ($\theta_{13}$) at low energy scale, consistent with the latest cosmological upper bound on the sum of three absolute neutrino mass eigenvalues, $\sum |m_{i}|
 < 0.12$ eV, is mainly addressed in the present study. Two cases of neutrino mass hierarchical models namely normal hierarchy (NH) and  inverted hierarchy (IH)  are considered when we take input mass eigenvalues at very high energy seesaw scale.
\newline

 A brief description of an exact form of golden ratio mixing matrix ($U_{GR}$) is given by \cite{king2017unified},

\begin{equation}
U_{GR}=\left(\begin{array}{ccc}
\frac{\phi}{\sqrt{2+\phi}}&\frac{1}{\sqrt{2+\phi}}&0\\
-\frac{1}{\sqrt{4+2\phi}}&\frac{\phi}{\sqrt{4+2\phi}}&\frac{1}{\sqrt{2}}\\
\frac{1}{\sqrt{4+2\phi}}&-\frac{\phi}{\sqrt{4+2\phi}}&\frac{1}{\sqrt{2}}
\end{array}\right)
\label{eq1}
\end{equation}
where $\phi$ has following properties : $$\phi=\phi^{2}-1=1+\frac{1}{\phi}=\frac{1+\sqrt{5}}{2}\approx 1.62$$
and
$$\frac{1}{\phi^{2}}=\frac{1-\frac{1}{\sqrt{5}}}{1+\frac{1}{\sqrt{5}}}\approx 0.382. $$ 
 It also predicts $\sin\theta_{13}=0$, $\sin\theta_{23}=\frac{1}{\sqrt{2}}$ and $\tan\theta_{12}=\frac{1}{\phi}$, leading to :
$$ \theta_{12}=\tan^{-1}\left(\frac{1}{\phi}\right)=31.72 ^{0}$$
and golden ratio is sometimes enforced by $A_{5}$ \cite{everett2009icosahedral}.
The $U_{GR}$ is a special case for the $\mu-\tau$ symmetric mass matrix,
$$ 
M_{\nu}=\left(\begin{array}{ccc}
D&A&\pm A\\
A&B&C\\
\pm A&C&B
\end{array}\right)
$$
where $$ \tan 2\theta_{12}=\frac{2\sqrt{2}A}{B\pm C-D}.$$
For case $B\pm C-D=A$, the mixing matrix goes to the tri-bimaximal mixing matrix ($U_{TBM}$) with $\tan2\theta_{12}=2\sqrt{2}$, and for case $B\pm C-D=\sqrt{2}A$, the mixing matrix goes to $U_{GR}$ with  $\tan2\theta_{12}=2$ \cite{fukuyama2017twenty,koide2004universal}. When D=0, the structure of the mass matrix predicts $\tan^{2}\theta_{12}=\frac{m_{1}}{m_{2}}\approx 0.382$, and $m_{1}$ and $m_{2}$ are two neutrino mass eigenvalues. To check the validity of GR neutrino mixings at high energy scale, we consider a large value of tan$\beta>60$ in order to include large effects of radiative corrections in the calculation of neutrino masses, mixing angles and to satisfy the latest cosmological upper bound $\sum |m_{i}|<0.12$ eV \cite{lorenz2021reconstruction, choudhury2020updated} in both normal and inverted hierarchical mass models.
\newline 

The paper is organised as follows. In section 2, we briefly outline the main points on renormalisation group analysis for neutrino oscillation parameters with phases. In section 3, we present numerical analysis of RGEs for GR neutrino mixing matrix. In section 4, we give results and discussion. In section 5, we give summary and conclusion.
\section{Renormalisation group analysis for neutrino oscillation parameters with phases}
We briefly present the main formalism for the evolution of neutrino oscillation parameters \cite{ma2002all,antusch2005charged,antusch2002neutrino} from high energy seesaw scale to low energy scale through the RGEs with CP violating phase in the MSSM, including appropriate SUSY threshold corrections. The neutrino masses can be described by lowest-dimensional neutrino mass operator compatible with the gauge symmetries of the SM. This operator reads in the SM \cite{antusch2003running}
\begin{equation}
\mathcal{L}_{K}^{SM}=-\frac{1}{4}K_{gf}\bar{l}_{Lc}^{{C}^{g}}\varepsilon^{cd}\phi_{d}l_{Lb}^{f}\varepsilon^{ba}\phi_{a}+h.c. ,
\label{eq2}
\end{equation} 
and in its minimal supersymmetric extension, the MSSM
\begin{equation}
\mathcal{L}_{K}^{MSSM}=-\frac{1}{4}K_{gf}\mathbb{l}_{c}^{g}\varepsilon^{cd}\mathbb{h}_{d}^{(2)}\mathbb{l}_{b}^{f}\varepsilon^{ba}\mathbb{h}_{a}^{(2)}|_{\theta\theta}+h.c. ,
\label{eq3}
\end{equation}
where $l_{L}^{C}$ is the charge conjugate of a lepton doublet. $\varepsilon$ is the totally antisymmetric tensor in 2 dimensions, and a, b, c, d $\in$ \{1,2\} are $SU(2)_{L}$ indices. The double-stroke letters $\mathbb{l}$ and $\mathbb{h}$ denote lepton doublets and the up-type Higgs superfield in the MSSM. The coefficients $K_{gf}$ are of mass dimension -1 and related to the Majorana neutrino mass matrix as $M_{\nu}=K\langle H \rangle^{2}$, where $\langle H \rangle=174\text{GeV}$ is the vacuum expectation value of Higgs field ($v_{0}$).

The most plausible explanation for neutrino mass is given by see-saw mechanism \cite{davidson1987universal}.
The neutrino mass matrix $m_{LL}$(t) which is generally obtained from see-saw mechanism, is expressible in terms of K(t), the coefficient of the dimension five neutrino mass operator in the scale-dependent manner, $t=\ln(\mu/1\text{GeV})$,
\begin{equation}
m_{LL}(t)=v_{u}^{2}(t)K(t) 
\label{eq4}
\end{equation}
where the vacuum expectation value (VEV) is $v_{u}=v_{o}\sin\beta$ and $v_{o}$=174 GeV in the minimal supersymmetric standard model (MSSM).
After diagonalization of K(t), the above eq.(\ref{eq4}) can be written in terms of mass eigenvalues as follows \cite{chankowski2000fixed}
\begin{equation}
m_{i}(t)=v_{u}^{2}(t)K_{i}(t); i=1,2,3.
\label{eq5}
\end{equation}
This expression can be simplified as
\begin{equation}
\frac{d}{dt}m_{i}(t)=m_{i}(t)\left(\frac{1}{K_{i}(t)}\frac{d}{dt} K_{i}(t)+\frac{2}{v_{u}(t)}\frac{d}{dt} v_{u}(t)\right).
\label{eq6}
\end{equation}
 Now, considering the phases in neutrino mixing matrix, we parameterize the Pontecorvo-Maki-Nakagawa-Sakata (PMNS) matrix as,
\begin{align}
U_{PMNS}=&\left(\begin{array}{ccc}
c_{13}c_{12}&c_{13}s_{12}&s_{13}e^{-i\delta}\\
-c_{23}s_{12}-c_{12}s_{13}s_{23}e^{-i\delta}&c_{12}c_{23}-s_{12}s_{13}s_{23}e^{-i\delta}&c_{13}s_{23}\\
s_{12}s_{23}-c_{12}s_{13}c_{23}e^{-i\delta}&-c_{12}s_{23}-c_{23}s_{13}s_{12}e^{-i\delta}&c_{13}c_{23}
\end{array}\right)\nonumber \\
\times & \left(\begin{array}{ccc}
e^{i\alpha_{1}}&0&0\\
0&e^{i\alpha_{2}}&0\\
0&0&1
\end{array}\right)
\label{eq7}
\end{align} 
where $s_{ij}=\sin\theta_{ij}$, $c_{ij}=\cos\theta_{ij}$, $\delta$= Dirac phase, $\alpha_{1}$=first Majorana phase, $\alpha_{2}$=second Majorana phase. Here, the three mixing angles are defined as $\tan\theta_{12}=\frac{|U_{e2}|}{|U_{e1}|}$, $\tan\theta_{23}=\frac{|U_{\mu 3}|}{|U_{\tau 3}|}$ and $\sin\theta_{13}=|U_{e3}|$.

The RGEs for $K_{i}(t)$ and $v_{u}(t)$ in the basis where charged lepton mass matrix is diagonal, for one-loop order in MSSM, in the energy range from $M_{R}$ to $M_{SUSY}$, are given by \cite{babu1993renormalization,casas2000nearly,arason1992renormalization}
 \begin{equation}
\frac{1}{K_{i}(t)}\frac{d}{dt} K_{i}(t)=\frac{1}{16\pi^{2}}\sum_{f=e,\mu,\tau}\left[-\frac{6}{5}g_{1}^{2}-6g_{2}^{2}+6h_{t}^{2}+2h_{f}^{2}U_{fi}^{2}\right]
\label{eq8}
\end{equation}
and
\begin{equation}
\frac{1}{v_{u}(t)}\frac{d}{dt} v_{u}(t)=\frac{1}{16\pi^{2}}\left[\frac{3}{20}g_{1}^{2}+\frac{3}{4}g_{2}^{2}-3h_{t}^{2}\right].
\label{eq9}
\end{equation}
The RGEs for $K_{i}$ and $v_{0}$ in the basis where the charged lepton mass matrix is diagonal, for one-loop order in SM, in the energy range from $M_{SUSY}$ to $M_{Z}$, are given by \cite{babu1993renormalization,casas2000nearly,arason1992renormalization}
 \begin{equation}
\frac{1}{K_{i}(t)}\frac{d}{dt} K_{i}(t)=\frac{1}{16\pi^{2}}\sum_{f=e,\mu,\tau}\left[-3g_{2}^{2}+2\lambda+6h_{t}^{2}+h_{\tau}^{2}+h_{f}^{2}U_{fi}^{2}\right]
\label{eq10}
\end{equation}
and
\begin{equation}
\frac{1}{v_{0}(t)}\frac{d}{dt} v_{0}(t)=\frac{1}{16\pi^{2}}\left[\frac{9}{20}g_{1}^{2}+\frac{9}{4}g_{2}^{2}-3h_{t}^{2}-3h_{b}^{2}-h_{\tau}^{2}\right],
\label{eq11}
\end{equation}
where $g_{1}, g_{2}$ are gauge couplings, and $h_{t}, h_{b}, h_{\tau}$ and $\lambda$  are top-quark, bottom -quark,  tau -lepton Yukawa couplings and SM quartic Higgs coupling respectively. As VEV can affect mass terms in the RGEs, we have two possible set of RGEs of neutrino masses where one is scale-dependent VEV and other is scale-independent VEV. The RGEs of neutrino mass eigenvalues for both scale dependent VEV and scale independent VEV can be written as \cite{agarwalla2007neutrino,antusch2003running}
\begin{equation}
\frac{d}{dt}m_{i}=-2F_{\tau}\left(P_{i}+Q_{i}\right)m_{i}-F_{u}m_{i},(i=1,2,3).
\label{eq12}
\end{equation}

where, $$ P_{1}=s_{12}^{2}s_{23}^{2}, P_{2}=c_{12}^{2}s_{23}^{2}, P_{3}=c_{13}^{2}c_{23}^2,$$
$$ Q_{1}=-\frac{1}{2}s_{13}\sin2\theta_{12}\sin2\theta_{23}\cos\delta+s_{13}^{2}c_{12}^{2}c_{23}^{2},$$
$$ Q_{2}=\frac{1}{2}s_{13}\sin2\theta_{12}\sin2\theta_{23}\cos\delta+s_{13}^{2}s_{12}^{2}c_{23}^{2},$$
$$ Q_{3}=0.$$
For scale-dependent VEV in the case of MSSM with $\mu \geq m_{s}$,
$$ F_{\tau}=-\frac{h_\tau^{2}}{16\pi^{2}cos^{2}\beta},\; F_{u}=\frac{1}{16\pi^{2}}\left(\frac{9}{10}g_{1}^{2}+\frac{9}{2}g_{2}^{2}\right)$$
but, for SM case with $\mu\geq m_{s}$,

$$ F_{\tau}=\frac{3h_\tau^{2}}{32\pi^{2}},\; F_{u}=\frac{1}{16\pi^{2}}\left(-\frac{9}{10}g_{1}^{2}-\frac{3}{2}g_{2}^{2}+6h_{b}^{2}-2\lambda\right).$$

For scale-independent VEV in the case of MSSM with $\mu \geq m_{s}$,
$$ F_{\tau}=-\frac{h_\tau^{2}}{16\pi^{2}\cos^{2}\beta},\; F_{u}=\frac{1}{16\pi^{2}}\left(\frac{6}{5}g_{1}^{2}+6g_{2}^{2}-6\frac{h_{t}^{2}}{\sin\beta^{2}}\right)$$
but, for SM case with $\mu\geq m_{s}$,

$$ F_{\tau}=\frac{3h_\tau^{2}}{32\pi^{2}},\; F_{u}=\frac{1}{16\pi^{2}}\left(3g_{2}^{2}-2\lambda-6h_{t}^{2}\right).$$

 For the present analysis, we adopt usual sign convention $|m_{2}|>|m_{1}|$ and we shall use RGEs of scale dependent VEV, which are different from RGEs of scale independent VEV in the expression of $F_{u}$ involved in the equations. The general case of fermion masses which decrease with the increase in energy scale is consistent with that of neutrino masses in the running of RGEs with VEV \cite{das2005numerical,singh2001effects}. The RGEs of gauge and Yukawa couplings with and without SUSY, are given in Appendix-A. The corresponding RGEs for three mixing angles and three phases are given in Appendix-B.

\section{Numerical analysis of RGEs for GR neutrino mixing matrix}
For a complete numerical analysis of the RGEs given in the above section, we follow here two consecutive steps: (i) bottom-up running \cite{parida1998low} in the first step and then (ii) top-down running\cite{king2001inverted} in the next step.

 In the first step (i), the running of the RGEs for the third family Yukawa couplings ($h_{t},h_{b},h_{\tau}$) and three gauge couplings ($g_{1},g_{2},g_{3}$) is carried out from top quark mass scale, $t_{0}=\ln(m_{t}/\text1{GeV})$ at low energy end to high energy scale $M_{R}$ via SUSY breaking scale $m_s$ where ($m_t< m_s< M_R$).

  At the transition point from SM to MSSM, the appropriate matching conditions without threshold corrections are given as follows \cite{singh2018stability},
\begin{equation}
g_{i}(SUSY)=g_{i}(SM),
\end{equation}
\begin{equation}
h_{t}(SUSY)= \frac{h_{t}(SM)}{\sin\beta},
\end{equation}
\begin{equation}
h_{b}(SUSY)=\frac{h_{b}(SM)}{\cos\beta},
\end{equation}
\begin{equation}
h_{\tau}(SUSY)=\frac{h_{\tau}(SM)}{\cos\beta},
\end{equation}
where $\tan\beta$=$\frac{v_{u}}{ v_{d}}$ such that $v_{u}=v_{o}\sin\beta$, $v_{d}=v_{o}\cos\beta$
and $v_{o}=174$ GeV is the VEV of the Higgs field \cite{alekhin2012top}.

For large value of $\tan\beta$, there should be SUSY threshold corrections which would lead to the modification of down-type quark and charged-lepton Yukawa coupling constants at the matching condition of SUSY breaking scale ($m_{s}$) as follows \cite{antusch2013running,zhang2016viability,antusch2008quark},
\begin{equation}
h_{t}(SUSY) \simeq \frac{h_{t}(SM)}{\sin\bar{\beta}},
\end{equation}
\begin{equation}
h_{b}(SUSY) \simeq \left(\frac{1}{1+\bar{\eta_{b}}}\right)\frac{h_{b}(SM)}{\cos \bar{\beta}},
\end{equation}
\begin{equation}
h_{\tau}(SUSY)\simeq \frac{h_{\tau}(SM)}{\cos \bar{\beta}},
\end{equation}
where $\bar{\eta_{b}}$ is a free parameter that describes the SUSY threshold corrections, cos$\bar{\beta}=(1+\eta_{l}^{'})\cos\beta$ in the redefinition of $\beta\rightarrow\bar{\beta}$ and $\eta_{l}^{'}$ is a leptonic SUSY threshold correction parameter which is typically very small. Neglecting the effect of the leptonic threshold correction parameters in our parametrisation, it would simply mean that tan$\bar{\beta}$=tan$\beta$.

 The latest experimental input values for physical fermion masses, gauge couplings and Weinberg mixing angle at electroweak scale ($m_{Z}$)\cite{pdg} are given in Table~\ref{t1}.
\begin{table}[H]
\begin{center}
\begin{tabular}{|c|c|c|} \hline
Mass in GeV&Coupling constant&Weinberg mixing angle \\ \hline
$m_{z}$($m_{Z}$)=91.1876&$\alpha_{em}^{-1}$($m_{Z}$)=127.952&$sin^{2}\theta_{W}(m_{Z})$=0.23121 \\
$m_{t}$($m_{t}$)=172.76&$\alpha_{s}(m_{Z}$)=0.1179& \\
$m_{b}$($m_{b}$)=4.18&& \\
$m_{\tau}$($m_{\tau}$)=1.7768&& \\  \hline
\end{tabular}
\end{center}
\hfill
\caption{\footnotesize Low energy experimental values of fermion masses, gauge coupling constants and Weinberg mixing angle.}
\label{t1}
\end{table}

 The three gauge couplings, $\alpha_{1}(m_{Z})=0.016943$ , $\alpha_{2}(m_{Z})=0.033802$ and $\alpha_{3}(m_{Z})=0.1179$ at low energy scale ($m{_Z}$), are calculated by using latest PDG data given in Table 1, and SM matching relations
\begin{equation}
 \frac{1}{\alpha_{em}(m_{Z})}=\frac{3}{5}\frac{1}{\alpha_{1}(m_{Z})}+\frac{1}{\alpha_{2}(m_{Z})}; 
\end{equation}
\begin{equation}
sin^{2}\theta_{W}(m_{Z})=\frac{\alpha_{em}(m_{Z})}{\alpha_{2}(m_{Z})}.
\end{equation}

 In terms of the normalized coupling constant $(g_{i})$, $\alpha_{i}$ can be expressed as $g_{i}=\sqrt{4\pi\alpha_{i}}$, where $i=1,2,3$ and it represents electromagnetic, weak and strong coupling constants respectively. We adopt the standard procedure to get the values of gauge couplings at top-quark mass scale from the experimental measurements at $m_{Z}$, using one-loop RGEs for simplicity, assuming the existence of one-light Higgs doublet and five quark flavours below $m_{t}$ scale\cite{parida1998low, singh2001effects}. 
 
 The evolution equation of gauge coupling constants of one loop for energy range $m_{Z}\leq\mu\leq m_{t}$ in SM is given by
 \begin{equation}
 \frac{1}{\alpha_{i}(\mu)}=\frac{1}{\alpha_{i}(m_{Z})}-\frac{b_{i}}{2\pi}ln\left(\frac{\mu}{m_{Z}}\right)
 \end{equation}
  where 
 $$ b_{i}=\left(\frac{53}{10}, -\frac{1}{2}, -4.0\right).$$
 
  Similarly, the Yukawa couplings are also evaluated at top-quark mass scale using QCD-QED rescaling factors ($\eta_{i}$) in the standard fashion\cite{singh2001effects}, which are given by following relations.

\begin{equation}
h_{t}(m_{t})=\frac{ m_{t}(m_{t})}{v_{0}};
\end{equation}

\begin{equation}
h_{b}(m_{t})=\frac{m_{b}( m_{b})}{v_{0}\eta_{b}}; 
\end{equation}
\begin{equation}
h_{\tau}(m_{t})=\frac{m_{\tau}( m_{\tau})}{v_{0}\eta_{\tau}}.
\end{equation}
The value of QCD-QED rescaling factors ($\eta_{i}$) and vacuum expectation ($v_{0}$) of Higgs field are given by $\eta_{b}=1.53$, $\eta_{\tau}=1.015$ and $v_{0}=174$ GeV respectively\cite{deshpande1994predictive,nimai1998third}.

In the second step (ii), the runnings of three neutrino mass eigenvalues ($m_{1},m_{2},m_{3}$),  three neutrino mixing angles ($s_{23}, s_{13}, s_{12}$) and three phases ($\delta, \alpha_{1}, \alpha_{2}$) are carried out, together with the running of gauge and Yukuwa couplings, from high energy seesaw scale $t_{R}$(=ln ($M_{R}$/1GeV)) to low energy scale $t_{o}(=\ln (m_{t}/\text{1GeV}))$ via SUSY breaking scale $t_{s}(=\ln (m_{s}/\text{1GeV}))$. In this case, we use the values of gauge and Yukawa couplings evaluated earlier at the scale $t_{R}$ from the first stage running of RGEs in (i). In principle, one can evaluate neutrino masses, mixing angles and phases\cite{murayama2002theory} at every point in the energy scale. 
\newline

  The present numerical analysis has six unknown arbitrary input values at high energy seesaw scale consisting of three neutrino masses and three phases. These input values would be suitably chosen so that we can get the desired low energy values of neutrino parameters and latest Planck cosmological upper bound on the sum of three neutrino masses. When we choose a set of mass eigenvalues, there are two possible cases of mass hierarchy, namely (i) normal hierarchy ($m_{3}>>m_{2}>m_{1}$) and (ii) inverted hierarchy ($m_{2}>m_{1}>>m_{3}$) as we generally set to $|m_{2}|>|m_{1}|$. 
 The three neutrino mixing angles ($s_{23},s_{13},s_{12}$) used at high energy seesaw scale ($M_{R}$), are given by the golden ratio mixing matrix, which are constant input values in all different high energy scales, while the three neutrino mass eigenvalues ($m_{1},m_{2},m_{3}$) and  phases ($\delta, \alpha_{1}, \alpha_{2}$) are suitably chosen input values which may give the desired values of neutrino oscillation parameters ($s_{23},s_{13},s_{12},m_{1},m_{2},m_{3}, \delta,\alpha_{1},\alpha_{2}$) at low energy scale after taking  radiative corrections. The main concern in our work is to satisfy the latest upper cosmological bound on the sum of absolute neutrino masses, $ \sum |m_{i}|<0.12$ eV \cite{lorenz2021reconstruction, choudhury2020updated} with the generation of reactor angle, $|U_{e3}|$ at low energy scale.

\section{Results and discussion}

For top-down running of RGEs from high to low energy scale, Table 2 represents the high scale input parameters of gauge and Yukawa coupling constants which are already evaluated in the bottom-up approach, for running the neutrino oscillation parameters. In the numerical analysis of neutrino oscillation parameters along with phases including  SUSY threshold corrections, there are four free parameters namely, $M_{R}$, $\tan\beta$, $\bar{\eta_{b}}$ and $m_{s}$ which may affect the values of coupling constants.  While taking a set of coupling constants at high energy scale ($M_{R}$), four possible cases are considered, where three of the four parameters are set to be fixed while other one is taken as variable. The values of the parameters are suitably chosen within the certain limit of ranges for checking the data of the output results for all cases. In addition to a particular set of coupling constants at high energy scale, we have nine neutrino oscillation parameters namely, three mass eigenvalues ($m_{1}, -m_{2}, m_{3}$), three neutrino mixing angles ($\theta_{23}, \theta_{12}, \theta_{13}$) and three  phases ($\delta,\alpha_{1}, \alpha_{2}$). The negative sign in mass eigenvalues are possible due to the absorption of two Majorana phases in the mass eigenvalues as diag($m_{1}e^{i\bar{\alpha_{1}}}, m_{2}e^{i\bar{\alpha_{2}}}, m_{3}$) where we consider $\bar{\alpha_{1}}=\alpha_{1}$ and $\bar{\alpha_{2}}=\alpha_{2}+\pi$. This negative sign in the mass eigenvalues may help to prevent from the possible singularity that may arise in the evolution of RGEs which has such $(m_{i}-m_{j})$ term in the denominator. Since we are considering GR neutrino mixing matrix, the three neutrino mixing angles at the high energy scale, are given by $s_{23}=0.70710, s_{13}=0$ and $ s_{12}=0.52573$ respectively, and these input values are the same in all cases. Since the reactor mixing angle ($\theta_{13}$) is exactly zero, we take it to have an extremely small non-zero value which would be able to solve the asymptotic function in the RGEs of Dirac phase. Now, the unknown arbitrary input values at high energy scale, are reduced to only six parameters i.e., three mass eigenvalues and three phases. These six arbitrary input values defined at high energy scale, should be suitably chosen so that the output results are compatible with the low energy experimental neutrino oscillation data\cite{mc,acero2022improved}, including latest cosmological upper bound.

 We have considered both normal and inverted hierarchical mass models for the numerical analysis. In the case of normal hierarchical mass model, all the low energy neutrino parameters are found to lie within $3\sigma$ range of NuFIT data\cite{mc} with $\sum |m_{i}|<0.12$ eV as shown in Tables \ref{t3}-\ref{t6}. We also check the case of inverted hierarchical mass model which fails to give the low energy neutrino oscillation parameters and $\sum |m_{i}|<0.12$ eV within the experimental bounds. We also study the radiative generation of $U_{e3}$ with initial conditions, $\Delta m_{21}^{2}$=0 at high energy scale, and a non-zero value of $m_{3}$, but it fails to give low energy experimental values of neutrino oscillation parameters. These results are not presented in the present work. We observe that in both cases of normal and inverted hierarchical models, all neutrino mass eigenvalues are slightly increased in magnitude with the decrease in energy scale, whereas the atmospheric mixing angle ($s_{23}$) and solar mixing angle ($s_{12}$) are slightly deviated from the mixing angles at high energy seesaw scale i.e $\theta_{23}>45^{0}$ for NH and $\theta_{23}<45^{0}$ for IH.

Our detailed numerical analysis shows that a larger value of $\tan\beta>$60 and high energy scale ($M_{R}$) are preferred in order to satisfy the latest cosmological upper bound on the sum of three absolute neutrino mass eigenvalues, $\sum |m_{i}|<0.12$ eV. This result requires an additional SUSY threshold free parameter $\bar{\eta_{b}}$ in the range from -0.6 to +0.6 \cite{antusch2013running,zhang2016viability} arising from the threshold corrections of heavy SUSY particles\cite{muramatsu2020susy,mohapatra2005threshold,gupta2015renormalization,hollik2014lifting}. It is found that all the neutrino oscillation parameters are consistent with low energy experimental data  for $\bar{\eta_{b}}$=0.01 at high scale $M_{R}=10^{15}$ GeV, large value of $tan\beta=68$, and $m_{s}=1$ TeV. It is also observed that the negative values of $\bar{\eta_{b}}$ for the larger values of $\tan\beta$, are not feasible  in calculating the values of coupling constants at high energy scale in the normal hierarchical model. As a result, we discard the negative values of $\bar{\eta_{b}}$ in our numerical analysis.
  
 The four possible ways of numerical analysis for $\sum |m_{i}|<0.12$ eV based on high energy scale ($M_{R}$), $\tan\beta$, SUSY breaking scale ($m_{s}$) and SUSY threshold parameter ($\bar{\eta_{b}}$), are studied as follows :
 
  (a) Case-I : Taking fixed input values  $M_{R}=10^{15}$ GeV, $\tan\beta$=68, $m_{s}$=1TeV,  we vary with  $\bar{\eta_{b}}=0.01, 0.2, 0.4, 0.6$ and this case is allowed only when $\bar{\eta_{b}}$=0.01. Results are presented in Table 3 and Fig.1(a). 
  
  (b) Case-II : Taking fixed values $\tan\beta$=68, $m_{s}$=1TeV, $\bar{\eta_{b}}$=0.01, we vary with $M_{R}=10^{12}\text{GeV},M_{R}=10^{13}\text{GeV},M_{R}=10^{14}\text{GeV},M_{R}=10^{15}\text{GeV}$ and this case is allowed only when $M_{R}=10^{15}$ GeV. Results are presented in Table 4 and Fig.1(b). 
   
  (c) Case-III : Taking fixed values $M_{R}=10^{15}$ GeV, $m_{s}$=1TeV, $\bar{\eta_{b}}$=0.01, we vary with  $ \tan\beta=38,48,58,68$ and this case is allowed only when $\tan\beta=68$. Results are presented in Table 5 and Fig.1(c).
   
  (d) Case-IV : Taking fixed values $M_{R}=10^{15}$ GeV, $\tan\beta$=68, $\bar{\eta_{b}}$=0.01, we vary with $ m_{s}=1\text{TeV},5\text{TeV},10\text{TeV},14\text{TeV}$ and this case is allowed only when $m_{s}$=1TeV. Results are presented in Table 6 and Fig.1(d).
    
     The required values of coupling constants for various cases are given in Table 2. The main numerical results of our analysis on neutrino oscillation parameters with three phases and SUSY threshold corrections, are given in Tables \ref{t3}-\ref{t6}. The numerical values in the upper halves of the tables represent the high energy scale input values of neutrino oscillation parameters and the numerical values in the lower halves of the Tables \ref{t3}-\ref{t6} represent the neutrino oscillation parameters at low energy scale. From numerical analysis of Table \ref{t3}, it indicates that a small value of $\bar{\eta_{b}}=0.01$ can accommodate the latest cosmological upper bound on the sum of three absolute neutrino mass eigenvalues $\sum |m_{i}|<0.12$ eV and hence, this particular value of $\bar{\eta_{b}}=0.01$ is fixed for the remaining three possible ways. From the numerical analysis of Table \ref{t4}-\ref{t6}, we also observe that the cosmological upper bound on the sum of three absolute neutrino mass eigenvalues $\sum |m_{i}|<0.12$ eV and the desired values of neutrino oscillation parameters at low energy scale can be achieved for the cases at $M_{R}=10^{15}$ GeV , $\tan\beta=68$, $\bar{\eta_{b}}=0.01$ and $m_{s}=1$ TeV  in the variation of these parameters. It is also observed that the input values of CP violating Dirac phase and two Majorana phases at high energy scale, have significant effects in the evolution of neutrino mixing angles at low energy scale. The best suitable high energy input values for these phase parameters are respectively found to be $\delta=175^{0}$, $\alpha_{1}=2^{0}$ and $\alpha_{2}=0.5^{0}$ for all possible cases in our analysis. The variation of $\sum |m_{i}|$ with $\bar{\eta_{b}}$, $M_{R}$, $\tan\beta$ and $m_{s}$ for (a) Case-I, (b) Case-II, (c) Case-III and (d) Case-IV are respectively shown in Fig: 1(a, b, c and d). We have also extended our analysis for other cases, where we fix $M_{R}=10^{15}$ GeV, $\tan\beta=68$, $m_{s}$=5 TeV, 10 TeV and 14 TeV respectively with various values of $\bar{\eta_{b}}=0.01, 0.2, 0.4, 0.6$. The  variations of $\sum |m_{i}|$ with $\bar{\eta_{b}}$ for each case is shown in Fig.2. Only case with $m_{s}=1$TeV falls within the acceptable region but all cases with higher $m_{s}$ are also acceptable if $\sum {|m_{i}|}< 0.23$ eV \cite{ade2016planck}. Hence, the case for inverted hierarchical model is not presented in this work as it fails to give latest Planck cosmological bound on the sum of three absolute neutrino mass eigenvalues,$\sum {|m_{i}|}< 0.12$ eV.

Other similar work in the literature \cite{zhang2016viability} emphasises the fact that if the Planck 2015 cosmological bound $\sum |m_{i}|<0.23$ eV\cite{ade2016planck} is taken into account, none of the three mixing patterns (BM, TBM, GR) can be identified as lepton mixing matrix below the seesaw threshold under radiative corrections. However, our present work shows the validity of the mixing pattern based on GR, which is consistent with the latest Planck 2021 cosmological bound $\sum |m_{i}|<0.12$ eV\cite{lorenz2021reconstruction, choudhury2020updated} for low SUSY breaking scale, $m_{s}=1$TeV in the normal hierarchy. As explained before, the numerical analysis in the present work is carried out with specific input parameters viz. $\tan\beta=68$, $M_{R}=10^{15}$GeV, $m_{s}=1$TeV and $\bar{\eta_{b}}=0.01$. For higher values of $1\text{TeV}\leq m_{s}\leq14\text{TeV}$ with other input parameters $\tan\beta=68$, $M_{R}=10^{15}$GeV and $0.01\leq\bar{\eta_{b}}\leq0.6$, the validity of GR is still acceptable if the cosmological bound is relaxed up to old 2015 Planck bound, $\sum |m_{i}|<0.23$ \cite{ade2016planck} as shown in Fig.2.

We consider the three phase parameters - two Majorana phases ($\alpha_{1}$, $\alpha_{2}$), one Dirac CP-violating phase ($\delta$) along with one free parameter $\bar{\eta_{b}}$ to determine the characteristics of SUSY threshold corrections at the matching scale. Further, the work in Ref.\cite{zhang2016viability} establishes an important analytical correlation $\Delta\theta_{23}\geq\theta_{13}\tan\theta_{12}^{0}$, where $\theta_{12}^{0}$ is the high scale input value, which induces a severe tension with the observed $\theta_{23}$ and leads to exclusion of both GR and TBM at $3\sigma$ level if the cosmological upper bound on the sum of the three absolute masses is taken into account. Although the above correlation is not explicitly shown in the present work, the result of our numerical analysis still agrees with it. However, our analysis shows the validity of GR under the most stringent latest Planck cosmological bound, $\sum m_{i}<0.12$eV at high energy seesaw scale $M_{R}=10^{15}$ GeV with larger value of $\tan\beta=68$ whose values are beyond the range of inputs assigned in Ref.\cite{zhang2016viability}.
  
\begin{table}[H]
\begin{center}
\begin{tabular}{|c|c|c|c|c|} \hline
(a) Case-I&$\bar{\eta_{b}}$=0.01&$\bar{\eta_{b}}$=0.2&$\bar{\eta_{b}}$=0.4&$\bar{\eta_{b}}$=0.6 \\ \hline
$g_{1}$&0.6686&0.67031&0.67068&0.67085 \\ 
$g_{2}$&0.7&0.70264&0.70325&0.70354\\ 
$g_{3}$&0.72562&0.72722&0.72761&0.72779  \\ 
$h_{t}$&0.96331&0.75129&0.70179&0.67866  \\
$h_{b}$&2.12322&0.652&0.445&0.34961 \\ 
$h_{\tau}$&2.63357&1.09202&0.91413&0.84265 \\
$\lambda$&0.49032&0.49032&0.49032&0.49032 \\ \hline
(b) Case-II&${M_{R}}=10^{12}$ GeV&${M_{R}}=10^{13}$ GeV&${M_{R}}=10^{14}$ GeV&${M_{R}}=10^{15}$ GeV \\
\hline
$g_{1}$&0.59652&0.61805&0.64197&0.6686 \\ 
$g_{2}$&0.68535&0.69036&0.69532&0.7\\ 
$g_{3}$&0.78147&0.76158&0.74305&0.72562  \\ 
$h_{t}$&0.8677&0.88399&0.9115&0.96331  \\
$h_{b}$&1.19924&1.33557&1.57371&2.12322  \\ 
$h_{\tau}$&1.26012&1.45802&1.80121&2.63357 \\ 
$\lambda$&0.49032&0.49032&0.49032&0.49032 \\ \hline
(c) Case-III&$\tan\beta$=38&$\tan\beta$=48&$\tan\beta$=58&$\tan\beta$=68 \\ \hline
$g_{1}$&0.67149&0.6712&0.67066&0.6686 \\ 
$g_{2}$&0.70408&0.70365&0.70286&0.7\\ 
$g_{3}$&0.72792&0.72767&0.72722&0.72562  \\ 
$h_{t}$&0.66369&0.69434&0.75274&0.96331  \\
$h_{b}$&0.26635&0.39168&0.62803& 2.12322 \\ 
$h_{\tau}$&0.32940&0.47995&0.75944&2.63357 \\
$\lambda$&0.49032&0.49032&0.49032&0.49032 \\ \hline
(d) Case-IV&$m_{s}$=1 TeV&$m_{s}$=5 TeV&$m_{s}$=10 TeV&$m_{s}$=14 TeV \\ \hline
$g_{1}$&0.6686&0.66194&0.65883&0.65749 \\ 
$g_{2}$&0.7&0.68679&0.68089&0.67839\\ 
$g_{3}$&0.72562&0.71030&0.70373&0.70097  \\ 
$h_{t}$&0.96331&0.79242&0.75688&0.74448  \\
$h_{b}$&2.12322&1.12484&0.99582&0.95473  \\ 
$h_{\tau}$&2.63357&1.46905&1.32587&1.28061 \\
$\lambda$&0.49032&0.47493&0.46891&0.46747 \\ \hline

\end{tabular} 
\end{center}
\hfill
\caption{\footnotesize High energy scale input values of gauge coupling ($g_{1},g_{2},g_{3}$), Yukawa coupling ($h_{t},h_{b},h_{\tau}$) and quartic coupling ($\lambda$) constants at various SUSY threshold free parameter ($\bar{\eta_{b}}$)((a) Case-I), high energy seesaw scale ($M_{R}$) ((b) Case-II), SUSY matching condition parameter ($\tan\beta$) ((c) Case-III) and SUSY breaking scale ($m_{s}$)((d) Case-IV) for four possible cases}
\label{t2}
\end{table}

\begin{table}[H]
\begin{center}
\begin{tabular}{|c|c|c|c|c|} \hline
Parameter&$\bar{\eta_{b}}$=0.01&$\bar{\eta_{b}}$=0.2&$\bar{\eta_{b}}$=0.4&$\bar{\eta_{b}}$=0.6 \\ \hline
$m_{1}^{0}$[eV]&0.02269&0.03408&0.03768&0.03945 \\ 
$m_{2}^{0}$[eV]&-0.02580&-0.03633& -0.03974&-0.04144\\ 
$m_{3}^{0}$[eV]&0.04881&0.05281&0.05481&0.05582  \\ 
$s_{23}^{0}$&0.70710&0.70710&0.70710&0.70710  \\
$s_{13}^{0}$&0.0001&0.0001&0.0001&0.0001  \\ 
$s_{12}^{0}$&0.52573&0.52573&0.52573&0.52573 \\
$\delta^{0}$[$/^{0}$]&175&175&175&175\\
$\alpha_{1}^{0}$[$/^{0}$]&2&2&2&2\\
$\alpha_{2}^{0}$[$/^{0}$]&0.5&0.5&0.5&0.5\\
$\Delta m_{21}^{2}[10^{-5}eV^{2}]$&15.08&15.84&15.94&16.09\\
$\Delta m_{31}^{2}[10^{-3}eV^{2}]$&1.86&1.62&1.58&1.55\\ 
$\sum |m_{i}|[eV]$&0.0973&0.12322&0.13223&0.13671 \\ \hline \hline
$m_{1}$[eV]&0.03038&0.04760&0.05302&0.05569 \\ 
$m_{2}$[eV]&-0.03158&-0.04839& -0.05371&-0.05636\\ 
$m_{3}$[eV]&0.05798&0.06883&0.07268&0.07457  \\ 
$s_{23}$&0.78946&0.77451&0.77285&0.77220  \\
$s_{13}$&0.14289&0.14319&0.14393&0.14424  \\ 
$s_{12}$&0.53576&0.54040&0.54342&0.54336 \\
$\delta$[$/^{0}$]&201.19&206.15&208.5&208.21\\
$\alpha_{1}$[$/^{0}$]&12.5&16.63&18.41&18.41\\
$\alpha_{2}$[$/^{0}$]&4.61&6.26&6.99&6.99\\
$\Delta m_{21}^{2}[10^{-5}eV^{2}]$&7.46&7.54&7.3&7.46\\
$\Delta m_{31}^{2}[10^{-3}eV^{2}]$&2.43&2.47&2.47&2.45\\ 
$\sum |m_{i}|[eV]$&0.11994&0.16482&0.17941&0.18662 \\ \hline
\end{tabular} 
\end{center}
\hfill
\caption{\footnotesize (a) Case-I: Effect of variations with SUSY threshold parameter $\bar{\eta_{b}}$ on radiative generation of low energy neutrino parameters in MSSM while running from $M_{R}=10^{15}$ GeV to low scale $m_{t}=172.76$ GeV through $m_{s}$=1TeV for a particular value of $\tan\beta$=68. Only case for $\bar{\eta_{b}}$=0.01 is allowed }
\label{t3}
\end{table}
\begin{table}[H]
\begin{center}
\begin{tabular}{|c|c|c|c|c|} \hline
Parameter&$M_{R}=10^{12}$GeV&$M_{R}=10^{13}$GeV&$M_{R}=10^{14}$GeV&$M_{R}=10^{15}$GeV \\ \hline
$m_{1}^{0}$[eV]&0.04235&0.03605&0.02985&0.02269 \\ 
$m_{2}^{0}$[eV]&-0.04436&-0.03854& -0.03254&-0.02580\\ 
$m_{3}^{0}$[eV]&0.06112&0.05652&0.05252&0.04881  \\ 
$s_{23}^{0}$&0.70710&0.70710&0.70710&0.70710  \\
$s_{13}^{0}$&0.0001&0.0001&0.0001&0.0001  \\ 
$s_{12}^{0}$&0.52573&0.52573&0.52573&0.52573 \\
$\delta^{0}$[$/^{0}$]&175&175&175&175\\
$\alpha_{1}^{0}$[$/^{0}$]&2&2&2&2\\
$\alpha_{2}^{0}$[$/^{0}$]&0.5&0.5&0.5&0.5\\
$\Delta m_{21}^{2}[10^{-5}eV^{2}]$&19.92&18.57&16.78&15.08\\
$\Delta m_{31}^{2}[10^{-3}eV^{2}]$&1.94&1.89&1.86&1.86\\ 
$\sum |m_{i}|[eV]$&0.14811&0.13111&0.11491&0.09730 \\ \hline \hline
$m_{1}$[eV]&0.05343&0.04667&0.03951&0.03038 \\ 
$m_{2}$[eV]&-0.05414&-0.04750& -0.04043&-0.03158\\ 
$m_{3}$[eV]&0.07276&0.06813&0.06357&0.05798  \\ 
$s_{23}$&0.77231&0.77532&0.78007&0.78946  \\
$s_{13}$&0.14320&0.14411&0.14450&0.14289  \\ 
$s_{12}$&0.54104&0.53932&0.53923&0.53576 \\
$\delta$[$/^{0}$]&208.72&206.89&206.6&201.19\\
$\alpha_{1}$[$/^{0}$]&17.11&15.96&15.66&12.5\\
$\alpha_{2}$[$/^{0}$]&6.45&5.98&5.87&4.61\\
$\Delta m_{21}^{2}[10^{-5}eV^{2}]$&7.59&7.74&7.28&7.46\\
$\Delta m_{31}^{2}[10^{-3}eV^{2}]$&2.43&2.46&2.47&2.43\\ 
$\sum |m_{i}|[eV]$&0.18033&0.16230&0.14351&0.11994 \\ \hline
\end{tabular} 
\end{center}
\hfill
\caption{\footnotesize (b) Case-II: Effect of variation with high energy seesaw scale $M_{R}$ on radiative generation of low energy neutrino parameters in MSSM while running from $M_{R}$ GeV to low scale $m_{t}=172.76$ GeV for a particular values of $\tan\beta$=68, $\bar{\eta_{b}}=0.01$ and $m_{s}$=1TeV. Only case for $M_{R}=10^{15}$ GeV is allowed.}
\label{t4}
\end{table}

\begin{table}[H]
\begin{center}
\begin{tabular}{|c|c|c|c|c|} \hline
Parameter&$\tan\beta$=68&$\tan\beta$=58&$\tan\beta$=48&$\tan\beta$=38 \\ \hline
$m_{1}^{0}$[eV]&0.02269&0.04752&0.06752&0.09082 \\ 
$m_{2}^{0}$[eV]&-0.02580&-0.04930& -0.06882&-0.09177\\ 
$m_{3}^{0}$[eV]&0.04881&0.06181&0.07781&0.09821  \\ 
$s_{23}^{0}$&0.70710&0.70710&0.70710&0.70710  \\
$s_{13}^{0}$&0.0001&0.0001&0.0001&0.0001  \\ 
$s_{12}^{0}$&0.52573&0.52573&0.52573&0.52573 \\
$\delta^{0}$[$/^{0}$]&175&175&175&175\\
$\alpha_{1}^{0}$[$/^{0}$]&2&2&2&2\\
$\alpha_{2}^{0}$[$/^{0}$]&0.5&0.5&0.5&0.5\\
$\Delta m_{21}^{2}[10^{-5}eV^{2}]$&15.08&17.23&17.72&17.34\\
$\Delta m_{31}^{2}[10^{-3}eV^{2}]$&1.86&1.56&1.49&1.39\\ 
$\sum |m_{i}|[eV]$&0.0973&0.15863&0.21415&0.2808 \\ \hline \hline
$m_{1}$[eV]&0.03038&0.06764&0.09749&0.13210 \\ 
$m_{2}$[eV]&-0.03158&-0.06820& -0.09788&-0.13239\\ 
$m_{3}$[eV]&0.05798&0.08432&0.10981&0.14108  \\ 
$s_{23}$&0.78946&0.77197&0.76960&0.76817  \\
$s_{13}$&0.14289&0.14857&0.14755&0.14530  \\ 
$s_{12}$&0.53576&0.54926&0.55597&0.55416 \\
$\delta$[$/^{0}$]&201.19&213.25&216.8&214.85\\
$\alpha_{1}$[$/^{0}$]&12.5&21.35&24.29&23.59\\
$\alpha_{2}$[$/^{0}$]&4.61&8.23&9.51&9.2\\
$\Delta m_{21}^{2}[10^{-5}eV^{2}]$&7.46&7.59&7.52&7.63\\
$\Delta m_{31}^{2}[10^{-3}eV^{2}]$&2.43&2.53&2.55&2.45\\ 
$\sum |m_{i}|[eV]$&0.11994&0.22016&0.30518&0.40557 \\ \hline
\end{tabular} 
\end{center}
\hfill
\caption{\footnotesize (c) Case-III: Effect of variation with the values of tan$\beta$ on radiative generation of low energy neutrino parameters in MSSM while running from $M_{R}=10^{15}$ GeV to low scale $m_{t}=172.76$ GeV for a particular value of $m_{s}$=1TeV and $\bar{\eta_{b}}=0.01$. Only case for tan$\beta$=68 is allowed.}
\label{t5}
\end{table}

\begin{table}[H]
\begin{center}
\begin{tabular}{|c|c|c|c|c|} \hline
Parameter&$m_{s}$=1TeV&$m_{s}$=5TeV&$m_{s}$=10TeV&$m_{s}$=14TeV \\ \hline
$m_{1}^{0}$[eV]&0.02269&0.03018&0.03194&0.03304 \\ 
$m_{2}^{0}$[eV]&-0.02580&-0.03271& -0.03436&-0.03546\\ 
$m_{3}^{0}$[eV]&0.04881&0.05131&0.05222&0.05322  \\ 
$s_{23}^{0}$&0.70710&0.70710&0.70710&0.70710  \\
$s_{13}^{0}$&0.0001&0.0001&0.0001&0.0001  \\ 
$s_{12}^{0}$&0.52573&0.52573&0.52573&0.52573 \\
$\delta^{0}$[$/^{0}$]&175&175&175&175\\
$\alpha_{1}^{0}$[$/^{0}$]&2&2&2&2\\
$\alpha_{2}^{0}$[$/^{0}$]&0.5&0.5&0.5&0.5\\
$\Delta m_{21}^{2}[10^{-5}eV^{2}]$&15.08&15.91&16.04&16.57\\
$\Delta m_{31}^{2}[10^{-3}eV^{2}]$&1.86&1.72&1.70&1.74\\ 
$\sum |m_{i}|[eV]$&0.0973&0.1142&0.11852&0.12172 \\ \hline \hline
$m_{1}$[eV]&0.03038&0.04109&0.04346&0.04492 \\ 
$m_{2}$[eV]&-0.03158&-0.04199& -0.0443&-0.04576\\ 
$m_{3}$[eV]&0.05798&0.06424&0.06575&0.06711  \\ 
$s_{23}$&0.78946&0.77816&0.77639&0.77600  \\
$s_{13}$&0.14289&0.14360&0.14308&0.14347  \\ 
$s_{12}$&0.53576&0.53853&0.53949&0.53964 \\
$\delta$[$/^{0}$]&201.19&204.94&205.84&205.92\\
$\alpha_{1}$[$/^{0}$]&12.5&15.29&15.99&16.11\\
$\alpha_{2}$[$/^{0}$]&4.61&5.71&6&6.04\\
$\Delta m_{21}^{2}[10^{-5}eV^{2}]$&7.46&7.48&7.35&7.58\\
$\Delta m_{31}^{2}[10^{-3}eV^{2}]$&2.43&2.43&2.43&2.48\\ 
$\sum |m_{i}|[eV]$&0.11994&0.14732&0.15351&0.15779 \\ \hline
\end{tabular} 
\end{center}
\hfill
\caption{\footnotesize (d) Case-IV: Effect of the variation with SUSY breaking scale $m_{s}$ on radiative generation of low energy neutrino parameters in MSSM while running from $M_{R}=10^{15}$ GeV to low scale $m_{t}=172.76$ GeV  for a particular value of $\tan\beta$=68 and $\bar{\eta_{b}}=0.01$. Only case for $m_{s}$=1TeV is allowed}
\label{t6}
\end{table}

\begin{figure}[!h]
     \centering
     \subfigure[]{
         \includegraphics[width=0.4\textwidth]{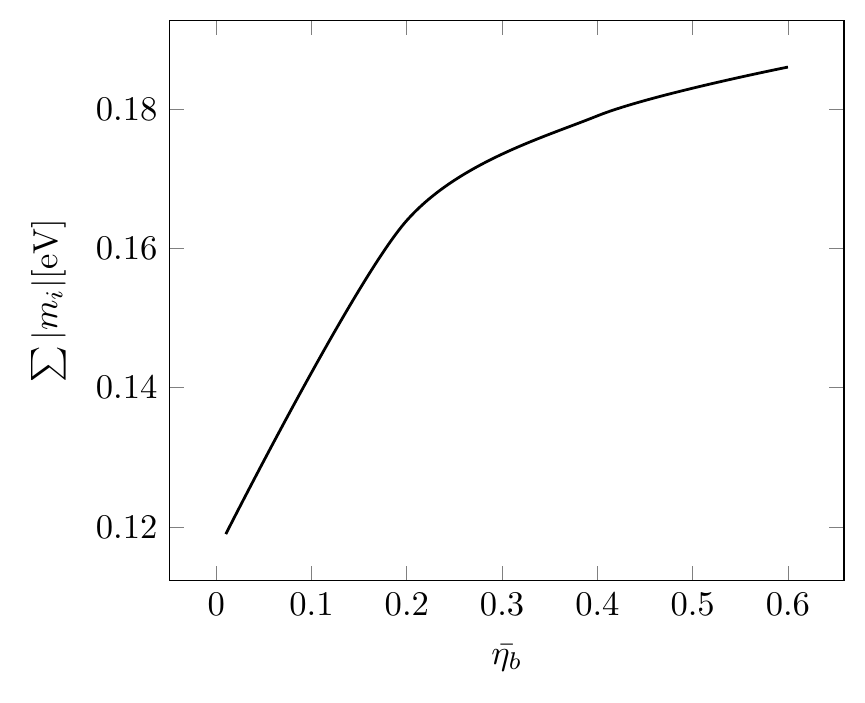}
          }
    \quad 
     \subfigure[]{
         \includegraphics[width=0.4\textwidth]{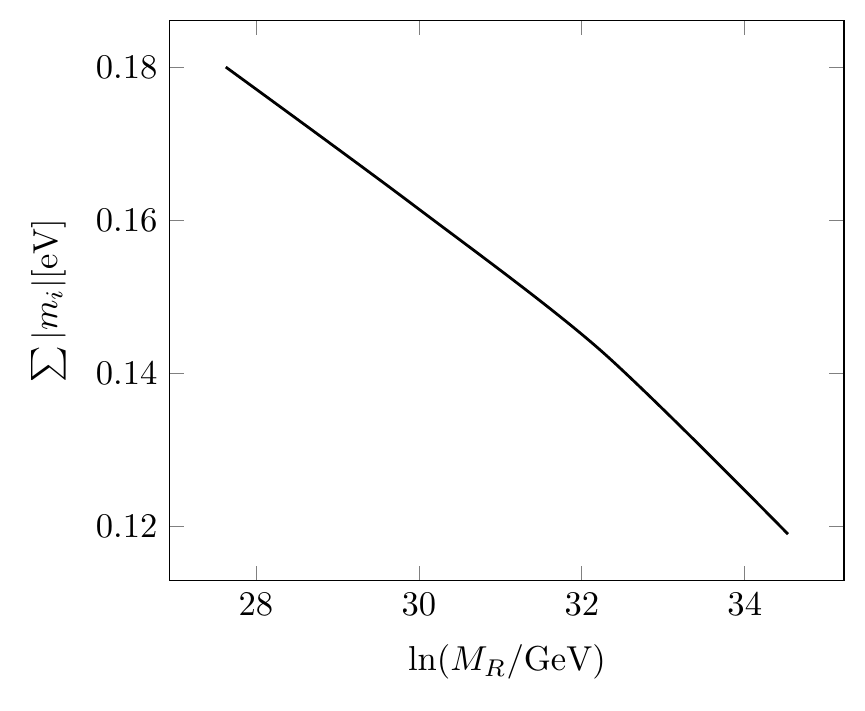}
        }
     \quad
     \subfigure[]{
         \includegraphics[width=0.4\textwidth]{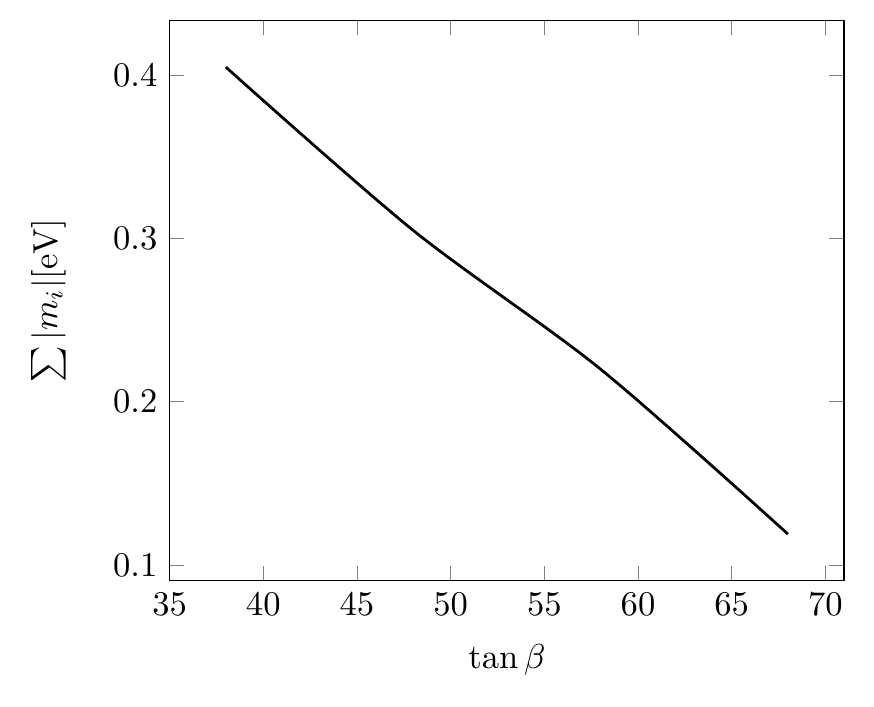}
         } 
     \quad
     \subfigure[]{
         \includegraphics[width=0.4\textwidth]{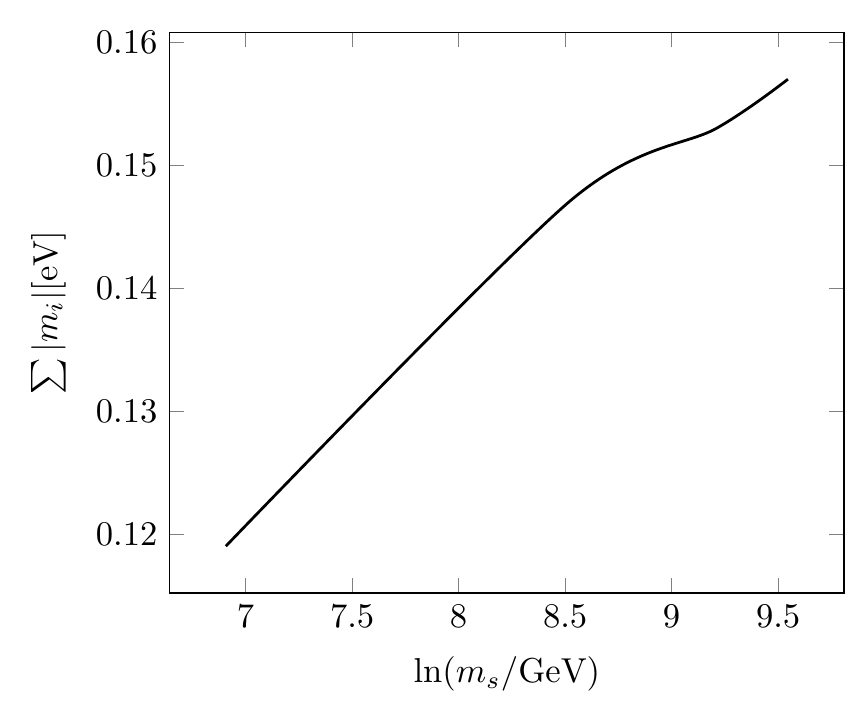}
         }
     \caption{Graphical presentation of the results given in Tables 3-6. (a) Case-I: Variation of $\sum |m_{i}|[\text{eV}]$ with $\bar{\eta_{b}}$ for $M_{R}=10^{15}\text{GeV}, \tan\beta=68, m_{s}=1\text{TeV}$, (b) Case-II: Variation of $\sum |m_{i}|[\text{eV}]$ with $M_{R}$ for $\bar{\eta_{b}}=0.01,\tan\beta=68, m_{s}=1\text{TeV}$, (c) Case-III: Variation of $\sum |m_{i}|[\text{eV}]$ with $\tan\beta$ for $\bar{\eta_{b}}=0.01, M_{R}=10^{15}\text{GeV}, m_{s}=1\text{TeV}$, (d) Case-IV: Variation of $\sum |m_{i}|[\text{eV}]$ with $m_{s}$ for $\bar{\eta_{b}}=0.01,\tan\beta=68, M_{R}=10^{15}\text{GeV}$}.
     \end{figure}
     
\begin{figure}
\centering
     \includegraphics[width=\textwidth]{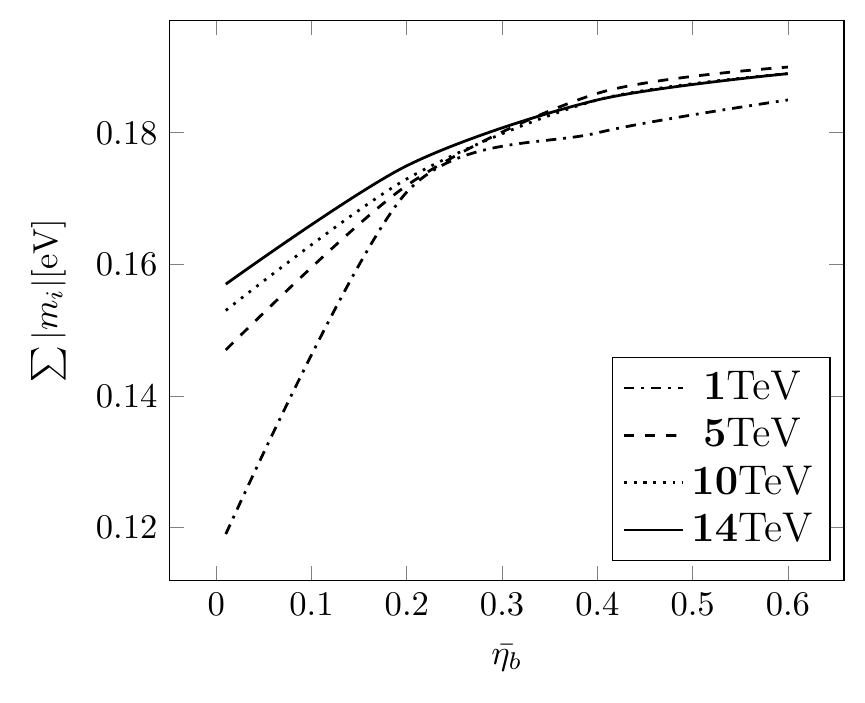}
\caption{Graphical presentation for variations of the sum of three absolute neutrino mass eigenvalues ($\sum |m_{i}|[\text{eV}]$) with the various values of SUSY threshold parameter ($\bar{\eta_{b}}=0.01, 0.2, 0.4, 0.6$) for different cases of SUSY breaking scale $m_{s}$=1 TeV, 5 TeV, 10 TeV and 14 TeV. Values of $M_{R}=10^{15}$GeV and $\tan\beta =68$ are taken. Higher upper bound $\sum |m_{i}|<0.23$ eV can accommodate at wide range of parameters, $\bar{\eta_{b}}$=(0.01-0.6) and $m_{s}$=(1TeV-14TeV)}.
   
\end{figure}

\pagebreak
\section{Summary and conclusion}
We summarise the main points of the present work related to the implication of the latest Planck data on the cosmological upper bound on the sum of the three absolute neutrino masses.

Our numerical analysis is based on the evolution of RGEs of neutrino oscillation parameters and three phases, including the effect of scale-dependent VEV and SUSY threshold corrections. We have first found out the most suitable value of SUSY threshold parameter $\bar{\eta_{b}}$ in the range from -0.6 to +0.6 \cite{antusch2013running,zhang2016viability}, which is compatible with the  low energy neutrino oscillation parameters and the most stringent cosmological upper bound, $\sum |m_{i}|<0.12$ eV \cite{lorenz2021reconstruction, choudhury2020updated}. It has been observed that the negative values of $\bar{\eta_{b}}$ are not feasible for large values of $\tan\beta$ and $M_{R}$ in the normal hierarchical model. The best fitted value of $\bar{\eta_{b}}$ that satisfies the cosmological upper bound $\sum |m_{i}|<0.12$ eV is found to be 0.01 when $M_{R}=10^{15}$GeV, $\tan\beta$=68 and $m_{s}$=1 TeV. 

The detailed numerical analysis shows that all neutrino mass eigenvalues as well as mixing angles are increased with the decrease in energy scale and it gives certain advantages on GR mixing matrix over the other two mixing matrices such as BM and TBM  in normal hierarchical mass model. It is also observed that the other neutrino mixing patterns such as BM and TBM mixings, do not satisfy the above cosmological upper bound in both normal hierarchical and inverted hierarchical models. The low energy neutrino oscillation parameters along with the latest Planck cosmological upper bound on sum of three absolute mass eigenvalues $\sum |m_{i}|<0.12$ eV can be achieved through radiative corrections under RGEs using GR neutrino mixing matrix defined at high energy seesaw scale for the  input values $M_{R}=10^{15}$ GeV, $\tan\beta$=68, $m_{s}$=1 TeV and $\bar{\eta_{b}}=0.01$. However, it is observed that a larger upper bound of $\sum |m_{i}|<0.23$ eV \cite{ade2016planck} can accommodate at wide range of parameters: $\bar{\eta_{b}}$=(0.01-0.6) and $m_{s}$=(1TeV-14TeV) as evident from Fig.2.

The analysis of inverted hierarchical neutrino mass model shows that the model does not accommodate the latest cosmological upper bound but it still conforms with earlier 2015 Planck bound $\sum |m_{i}| < 0.23$ eV \cite{ade2016planck}. Further analysis for TBM case \cite{chen2019cp,chen2019c}, considering the effect of  CP violating phases and SUSY threshold corrections, will be reported in future communication. 

To conclude, the present investigation indicates the sensitivity of the value of $\sum |m_{i}|$ on the origin of neutrino masses and mixing angles. 
It is relevant in the context of the information related to the absolute neutrino masses that has been continuously updating with recent Planck data on the cosmological upper bound on the sum of three absolute neutrino masses $\sum |m_{i}|<0.12$ eV. Neutrino mass model if any, is bound to be consistent with these upper bounds on absolute neutrino masses. While the existence of supersymmetric particles has been continuously ruling out in LHC, the supersymmetric breaking scale ($m_{s}$) still remains as an unknown parameter. We assume that the $m_{s}$ scale may lie somewhere in between 1 TeV and 14 TeV within the scope of LHC, and the present work is thus confined to the implication of SUSY breaking scale. It is a continuation of our previous investigation\cite{wilina2022deviations,devi2022effects,singh2018stability} on neutrino masses and mixings with varying SUSY breaking scale in the running of RGEs in both normal and inverted hierarchical neutrino mass models. 

The focus of the present work is the question of the validity of GR neutrino mixing at high energy scale, with the variation of $m_{s}$ scale and other input parameters $\tan\beta$, $\bar{\eta_{b}}$ and $M_{R}$ scale. It has profound implications to apply on other aspects of RGEs analysis such as low energy magnification of neutrino mixings in quark-lepton unification hypothesis at high energy scale in SO(10) model\cite{agarwalla2007neutrino,srivastava2016predictions,abdussalam2021majorana}, radiative generation of reactor mixing angle and solar neutrino mass squared difference at low scale, and the question of radiative stability of neutrino mass models to discriminate between NH and IH models. These earlier good results may now be readdressed for further analysis at low energy scale, consistent with latest Planck data on cosmological upper bound on the sum of three absolute mass eigenvalues.
\section*{Acknowledgements}
One of the authors (YMS) would like to thank  Manipur University for granting Fellowship for Ph.D course.
\appendix
\section{}

The one-loop RGEs for Yukawa couplings in the MSSM in the range of mass scales $m_{s}\leq \mu \leq M_{R}$ are given by \cite{rajkhowa2006possible}
\begin{equation}
\frac{d}{dt}h_{t}=\frac{h_{t}}{16\pi^{2}}\left(6h_{t}^{2}+h_{b}^{2}-\sum_{i=1}^{3}c_{i}g_{i}^2\right)
\end{equation},
\begin{equation}
\frac{d}{dt}h_{b}=\frac{h_{b}}{16\pi^{2}}\left(6h_{b}^{2}+h_{\tau}^{2}+h_{t}^{2}-\sum_{i=1}^{3}c_{i}^{'}g_{i}^2\right),
\end{equation}
\begin{equation}
\frac{d}{dt}h_{\tau}=\frac{h_{\tau}}{16\pi^{2}}\left(4h_{\tau}^{2}+3h_{b}^{2}-\sum_{i=1}^{3}c_{i}^{''}g_{i}^2\right)
\end{equation}
where, for SUSY case, $$ 
c_{i}=\left(\begin{array}{r}
\frac{13}{15},
3,
\frac{16}{3}
\end{array}\right),
c_{i}^{'}=\left(\begin{array}{r}
\frac{7}{15},
3,
\frac{16}{3}
\end{array}\right),
c_{i}^{''}=\left(\begin{array}{r}
\frac{9}{5},
3,
0
\end{array}\right).
$$
The Yukawa RGEs, for non-SUSY(SM) in the range of mass scales $m_{t}\leq \mu \leq m_{s}$,
\begin{equation}
\frac{d}{dt}h_{t}=\frac{h_{t}}{16\pi^{2}}\left(\frac{9}{2} h_{t}^{2}+\frac{3}{2} h_{b}^{2}+h_{\tau}^{2}-\sum_{i=1}^{3}c_{i}g_{i}^2\right),
\end{equation}
\begin{equation}
\frac{d}{dt}h_{b}=\frac{h_{b}}{16\pi^{2}}\left(\frac{9}{2} h_{b}^{2}+h_{\tau}^{2}+\frac{3}{2} h_{t}^{2}-\sum_{i=1}^{3}c_{i}^{'}g_{i}^2\right),
\end{equation}
\begin{equation}
\frac{d}{dt}h_{\tau}=\frac{h_{\tau}}{16\pi^{2}}\left(\frac{5}{2} h_{\tau}^{2}+3h_{b}^{2}+3h_{t}^2-\sum_{i=1}^{3}c_{i}^{''}g_{i}^2\right)
\end{equation}
where, for non-SUSY(SM) case,$$ 
c_{i}=\left(\begin{array}{r}
0.85,
2.25,
8.00
\end{array}\right),
c_{i}^{'}=\left(\begin{array}{r}
0.25,
2.25,
8.00
\end{array}\right),
c_{i}^{''}=\left(\begin{array}{r}
2.25,
2.25,
0.0
\end{array}\right).
$$
And one loop RGE for quartic coupling in SM is given by
\begin{equation}
\frac{d\lambda}{dt}=\frac{1}{16\pi^{2}}[\frac{9}{4}(\frac{3}{25}g_{1}^{4} +\frac{2}{5}g_{1}^{2}g_{2}^{2}+g_{2}^{4})$$ $$-(\frac{9}{5}g_{1}^{2}+9g_{2}^{2})\lambda+4Y_{2}(S)\lambda-4H(S)+12\lambda_{2}]
\end{equation}
where, 
$$ Y_{2}(S)=3h_{t}^{2}+3h_{b}^{2}+h_{\tau}^{2},$$ 
$$H(S)=3h_{t}^{4}+3h_{b}^{4}+h_{\tau}^{4},$$
$$\lambda=\frac{m_{h}^{2}}{v^{2}}$$ 
 $m_{h}$=Higgs mass, $v_{0}$=vacuum expectation value.

The two-loop RGEs for the gauge couplings are similarly expressed in the range of mass scales $m_{s}\leq\mu\leq M_{R}$ as \cite{singh2018stability}
\begin{equation}
\frac{d}{dt}g_{i}=\frac{g_{i}}{16\pi^{2}}\left[b_{i}g_{i}^{2}+\frac{1}{16\pi^{2}}\left(\sum_{j=1}^{3}b_{ij}g_{i}^{3}g_{j}^{2}-\sum_{j=t,b,\tau}a_{ij}g_{i}^{3}h_{j}^{2}\right)\right]
\end{equation}
where,  for SUSY case,
$$
b_{i}=\left(\begin{array}{r}
6.6,
1,
-3
\end{array}\right),
b_{ij}=\left(\begin{array}{ccc}
7.96&5.40&17.60\\
1.80&25.00&24.00\\
2.20&9.00&14.00
\end{array}\right),
a_{ij}=\left(\begin{array}{ccc}
5.2&2.8&3.6\\
6.0&6.0&2.0\\
4.0&4.0&0.0
\end{array}\right)
$$
and, for the non-SUSY(SM) in the range of mass scales $m_{t}\leq\mu\leq m_{s}$,
$$
b_{i}=\left(\begin{array}{r}
4.1,
-3.167,
-7
\end{array}\right),
b_{ij}=\left(\begin{array}{ccc}
3.98&2.70&8.8\\
0.90&5.83&12.0\\
1.10&4.50&-26.0
\end{array}\right),
a_{ij}=\left(\begin{array}{ccc}
0.85&0.5&0.5\\
1.50&1.5&0.5\\
2.0&2.0&0.0
\end{array}\right).
$$
\section{}
The RGEs of neutrino mixing angles and CP violating Dirac and Majorana phases are same for both scale-dependent VEV and scale independent VEV and are given by following equations\cite{agarwalla2007neutrino,antusch2003running} ,
\begin{align}
\frac{ds_{12}}{dt}=\frac{F_{\tau}c_{12}\sin2\theta_{12}s_{23}^{2}}{2(m_{2}^{2}-m_{1}^{2})}[m_{1}^{2}+m_{2}^{2}+2m_{1}m_{2}\cos(2\alpha_{2}-2\alpha_{1})],
\end{align}
\begin{align}
\frac{ds_{23}}{dt}=& \frac{F_{\tau} c_{23}\sin2\theta_{23}}{2(m_{3}^{2}-m_{2}^{2})}[c_{12}^{2}(m_{3}^{2}+m_{2}^{2}+2m_{3}m_{2}\cos2\alpha_{2}) \nonumber \\ 
&+s_{12}^{2}(m_{3}^{2}+m_{1}^{2}+2m_{3}m_{1}\cos2\alpha_{1})/(1+R)],
\end{align}
\begin{align}
\frac{ds_{13}}{dt}=& -\frac{F_{\tau} c_{13}\sin2\theta_{12}\sin2\theta_{23}m_{3}}{2(m_{3}^{2}-m_{1}^{2}}[m_{1}\cos(2\alpha_{1}-\delta)\nonumber \\
&-(1+R)m_{2}\cos(2\alpha_{2}-\delta)-Rm_{3}\cos\delta],
\end{align}
\begin{align}
\frac{d\delta}{dt}=& -\frac{F_{\tau}m_{3}\sin2\theta_{12}\sin2\theta_{23}}{2\theta_{13}(m_{3}^{2}-m_{1}^2)}\times[m_{1}\sin(2\alpha_{1}-\delta)-(1+R)m_{2}\sin(2\alpha_{2}-\delta)\nonumber \\
&+Rm_{3}\sin\delta]-2F_{\tau}[\frac{m_{1}m_{2}s_{23}^{2}\sin(2\alpha_{1}-2\alpha_{2})}{(m_{2}^{2}-m_{1}^{2})}+m_{3}s_{12}^{2}(\frac{m_{1}\cos2\theta_{23}\sin2\alpha_{1}}{(m_{3}^{2}-m_{1}^{2})}\nonumber \\&+\frac{m_{2}c_{23}^{2}\sin(2\delta-2\alpha_{2})}{(m_{3}^{2}-m_{2}^{2})}) 
+m_{3}c_{12}^{2}(\frac{m_{1}c_{23}^{2}\sin(2\delta-2\alpha_{1})}{(m_{3}^{2}-m_{1}^{2})}
\nonumber \\&+\frac{m_{2}\cos2\theta_{23}\sin2\alpha_{2}}{(m_{3}^{2}-m_{2}^{2})})],
\end{align}
\begin{align}
\frac{d\alpha_{1}}{dt}=&-2F_{\tau}[m_{3}\cos2\theta_{23}\frac{m_{1}s_{12}^{2}\sin\alpha_{1}+(1+R)m_{2}c_{12}^{2}\sin2\alpha_{2}}{m_{3}^{2}-m_{1}^{2}} \nonumber \\
&+\frac{m_{1}m_{2}c_{12}^{2}s_{23}^{2}\sin(2\alpha_{1}-\alpha_{2})}{m_{2}^{2}-m_{1}^{2}}],
\end{align}
\begin{align}
\frac{d\alpha_{2}}{dt}=&-2F_{\tau}[m_{3}\cos2\theta_{23}\frac{m_{1}s_{12}^{2}\sin\alpha_{1}+(1+R)m_{2}c_{12}^{2}\sin2\alpha_{2}}{m_{3}^{2}-m_{1}^{2}} \nonumber \\
&+\frac{m_{1}m_{2}s_{12}^{2}s_{23}^{2}\sin(2\alpha_{1}-\alpha_{2})}{m_{2}^{2}-m_{1}^{2}}],
\end{align}
where,
$$ R=\frac{(m_{2}^{2}-m_{1}^{2})}{(m_{3}^{2}-m_{2}^{2})}$$

\bibliographystyle{ieeetr}

\bibliography{validityofgr}

\end{document}